\documentclass[usenatbib,usegraphicx]{mn2e}
\pdfoutput=0
\usepackage{ifpdf}

\usepackage{amsmath}
\usepackage{graphicx}
\usepackage{color}
\usepackage{ulem}
\usepackage{fixltx2e}
\DeclareGraphicsExtensions{.eps,.pdf,.png,.jpg}

\usepackage[]{hyperref}

\hypersetup{
    colorlinks=true,
    linkcolor=black,
    citecolor=black,
    filecolor=black,
    urlcolor=black,
}

\begin{document}

    \ifpdf
    \else

\title[Magnus force in protoplanetary disks]{Curveballs in protoplanetary disks - the effect of the Magnus force on planet formation}

\author[J. C. Forbes]{John C. Forbes$^1$\textsuperscript{\thanks{E-mail: jcforbes@ucsc.edu}  }\\ 
$^1$Department of Astronomy \& Astrophysics, University of California, Santa Cruz, CA 95064 USA 
}

\maketitle

\begin{abstract}
Spinning planetesimals in a gaseous protoplanetary disk may experience a hydrodynamical force perpendicular to their relative velocities. We examine the effect this force has on the dynamics of these objects using analytical arguments based on a simple laminar disk model and numerical integrations of the equations of motion for individual grains. We focus in particular on meter-sized boulders traditionally expected to spiral in to the central star in as little as 100 years from 1 A.U. We find that there are plausible scenarios in which this force extends the lifetime of these solids in the disk by a factor of several. More importantly the velocities induced by the Magnus force can prevent the formation of planetesimals via gravitational instability in the inner disk if the size of the dust particles is larger than of order 10 cm. We find that the fastest growing linear modes of the streaming instability may still grow despite the diffusive effect of the Magnus force, but it remains to be seen how the Magnus force will alter the non-linear evolution of these instabilities. 
\end{abstract}
\begin{keywords}
Planetary systems -- protoplanetary disks, Planetary systems -- meteorites, meteors, meteoroids
\end{keywords}

\section{Introduction}

Planets are common, but the process by which they form may be complicated and is difficult to observe directly. Somehow the small dust grains present in the interstellar medium of galaxies must become $\ga1000$ km-sized planets. Many orders of magnitude in this `size ladder' are reasonably well-understood. Below $\sim 10$ cm, grains can grow by sticking together during collisions \citep[see][for an extensive review of this collisional process]{Blum2008}, while gravity is strong enough for objects larger than a few km that they can grow by gravitationally-focused collisions. Between these regimes lies the meter barrier, where solids are both too small for their self-gravity to be important, and are likely moving too fast relative to the gas and hence each other to stick upon colliding. To make matters worse, meter-sized particles experience particularly effective drag forces, causing them to spiral into the star at the center of the disk in an astronomically minuscule time of order 100 years from 1 A.U. It remains a mystery how exactly the solids which eventually become planets overcome this barrier.

Among the most influential ideas is direct gravitational collapse through the collective gravity of many grains. \citet{Goldreich1973} presented an early version of this idea, in which dust would settle out of the gas disk into a dynamically cold midplane where the dust disk would be subject to Toomre instability leading to direct collapse to $\sim 10$ km planetesimals. Although this scenario is unlikely owing to turbulence in the disk and Kelvin-Helmholz-like instabilities which increase the grain velocity dispersion \citep{weidenschilling1980}, a variety of physical mechanisms have been proposed to aid the grains in their gravitational collapse, including pressure traps associated with magneto rotational instability-induced turbulence \citep{fromang2005} and the streaming instability in which the drag on many individual particles has a large back-reaction on the gas when the density in solids is of order the density of the gas \citep[e.g.][]{goodman2000,Youdin2005}. A recent review by \citet{Johansen2014} details these processes and other aspects of planetesimal formation.

In this paper we explore the effects of another potentially important piece of physics. First described by \citet{newton1671}, the Magnus force is best-known for its effects on terrestrial sports \citep{Mehta1985} and aeronautics \citep{seifert2012}. Spinning objects moving with respect to a background fluid create an asymmetric wake in the fluid around them, which in turn alters the forces felt by the object. In addition to the drag force which opposes the motion of individual grains relative to the background gas, a component of the fluid's force on the object is perpendicular to the relative motion. When the circulation is caused primarily by the spin of the object, the force will also be perpendicular to the spin axis.

The addition of a force beyond the usually-considered drag and gravitational forces has the potential to change the mechanics of many of the scenarios commonly considered in planet formation, particularly for small objects where interaction with the gas is already known to be important. We can immediately guess that the force will tend to be diffusive, in that it will apply forces to individual particles isotropically, assuming that whatever process gives rise to the spin does not give rise to a preferred spin axis. All else being equal, we therefore expect that this effect will make planetesimal formation more difficult.

We will argue that there are plausible physical mechanisms that provide meter-sized solids with enough spin for the Magnus force to play an appreciable role in the dynamics of these objects. We will also argue based on our simulations of the dynamics of individual grains that in regimes where the Magnus force acts at all, disks of solid material are unlikely to be dynamically cold enough to collapse gravitationally in the inner disk unless the grains undergoing collapse are smaller than about 10 cm.

In section \ref{sec:classical}, we collect the basic ingredients of a laminar disk model and its interactions with individual particles. We detail our treatment of the Magnus force in section \ref{sec:magnus}, then explore the conditions under which it is relevant to the dynamics in section \ref{sec:analytic}. Focusing on one such regime, we carry out numerical integrations of the equations of motion including the Magnus effect in section \ref{sec:numerical}. In section \ref{sec:discussion} we analyze the results in a co-rotating reference frame, and show that the dynamics we observe in the integrations can be largely explained as the particles following force-free trajectories in this frame. We use this fact to map out the effects of the Magnus force in several planetesimal formation scenarios as a function of heliocentric location in the disk, particle size, and other relevant parameters. We summarize in section \ref{sec:summary}. 

\section{Classical scenario}
\label{sec:classical}
In this section we review the key issues regarding the flow of planetesimals in a proto-planetary disk. For simplicity we adopt a simple disk model following \citet{chiang2010}, wherein
\begin{eqnarray}
\label{eq:rhog}
\rho_g(r,z) &=& 2.79 \times 10^{-9} F\ r_{AU}^{-39/14} \mathrm{sech}^2\left( \frac{z}{h_g}\right)\ \mathrm{g}\ \mathrm{cm}^{-3} \\
h_g(r) &=& 0.022\ r_{AU}^{9/7}\ \mathrm{AU}\\
\lambda(r) &=& 0.5\ F^{-1} r_{AU}^{39/14}\ \mathrm{cm} \\
\eta(r) & = & 8\times 10^{-4} r_{AU}^{4/7} \\
c_s(r) &=& \sqrt{k_B T(r) / 2.2 m_H} \\
T(r) &=& 120\ r_{AU}^{-3/7}\ \mathrm{K}
\end{eqnarray}
respectively the gas density, gas scale height, gas mean free path, fractional velocity suppression of the gas by pressure support, gas sound speed, and gas temperature. We also define the Keplerian velocity $v_K = r\Omega_K = \sqrt{G M_*/r}$, for a central star of mass $M_*$. The cylindrical distance from the star is given in astronomical units as $r_{AU}$, while $k_B$ and $m_H$ represent Boltzmann's constant and the atomic mass of hydrogen. $F$ is a factor by which the disk's overall density may be scaled.

From the numbers above, we have almost all the information necessary to calculate the grain Reynolds number
\begin{equation}
Re = \frac{4 s v_{rel}}{\lambda c_s}
\end{equation}
where $s$ the radius of the grain. To find $\vec{v}_{rel}$, the relative velocity between the grain and the gas, we need to take into account the drag force felt by the grain. In the limit of ineffective drag, the magnitude of this velocity $v_{rel} \approx \eta v_K $, i.e. the planetesimal will move at nearly Keplerian velocity, whereas the gas will be slightly slower owing to its pressure support. Small grains that are well-coupled to the gas will have substantially smaller velocities, however.

The drag force experienced by individual grains operates in one of four regimes depending on the Reynolds number and the mean free path of gas molecules relative to the grain size. We take the dimensionless drag coefficient $C_D$, defined as the ratio of the drag force $F_D$ to $(1/2) \rho_g \pi s^2 v_{rel}^2$, to be
\begin{equation}
\label{eq:drag}
C_D = 
\begin{cases}
0.44  & \ \mbox{if}\ Re>800 \\
24 (Re)^{-0.6} & \ \mbox{if}\ 1<Re<800 \\
24 (Re)^{-1} & \ \mbox{if}\ Re<1\ \mbox{and}\ \lambda/s <4/9 \\
\frac{8}{3}  c_s/v_{rel} &\ \mbox{if}\ Re<1\ \mbox{and}\ \lambda/s > 4/9
\end{cases}
\end{equation}
following \citet{stepinski1996}. See also \citet{Weidenschilling1977, Garaud2004b, Loth2008}. In the high-$Re$ regime where inertial forces dominate viscous forces, the drag arises when the oncoming gas is deflected and slowed by the particle. In the high-$\lambda$ regime where the fluid equations break down, the drag arises via a slight difference in the momentum flux on either side of the particle, in part owing to the thermal motion of the gas particles. In between, viscous forces become important increasing the drag relative to the inertial regime. 

\begin{figure}
\centering
\hspace*{-.8cm}\includegraphics[width=8cm,clip=true,trim=1.5cm .3cm .8cm .3cm]{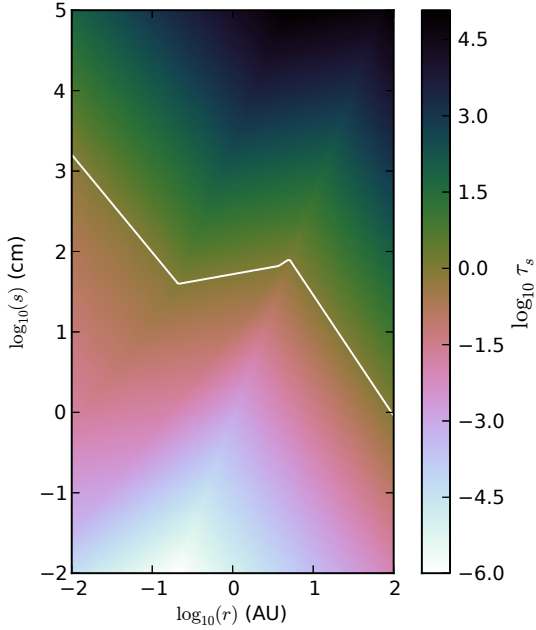}
\caption{The log of the non-dimensional stopping time of grains of various sizes at various locations in the disk. We show the contour where $\tau_s=1$, i.e. the dividing line between the grains being well-coupled to the gas via drag, and the grains largely ignoring the gas.}
\label{fig:taus}
\end{figure}

\begin{figure}
\centering
\hspace*{-1.1cm}\includegraphics[width=8cm,clip=true,trim=1.5cm .3cm .8cm .3cm]{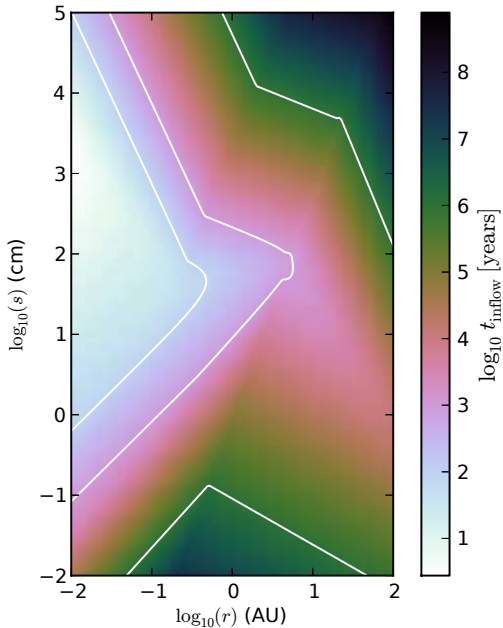}
\caption{The log of the inflow timescale $r/v_r$ given in years. We highlight 100 yr, 1000 yr, and 1 Myr. In the classical scenario, few if any particles can survive the full multi-Myr lifetime of the disk, and objects of order 1 meter in size can quickly spiral in through the disk to the star. }
\label{fig:tinflow}
\end{figure}

With the drag force in hand, we may now define the stopping time $t_s = m v_{rel}/F_d$ and its dimensionless version $\tau_s = t_s \Omega_K$. The velocity of the particle relative to Keplerian may be approximated following \citet{chiang2010},
\begin{equation}
\label{eq:vr}
v_r \approx -2\eta\Omega_K r\frac{\tau_s}{1+\tau_s^2}
\end{equation}
and
\begin{equation}
\label{eq:vphiRel}
v_\phi - v_K \approx \frac{-\eta\Omega_K r}{1+\tau_s^2}
\end{equation}
See also \citet{Weidenschilling1977, Youdin2010}. The particle has a tangential velocity relative to the gas of $v_\phi - (1-\eta)v_K = \eta v_K + (v_\phi-v_K)$, yielding a total relative velocity magnitude
\begin{equation}
\label{eq:vrel}
v_{rel} \approx \eta\Omega_K r \frac{\tau_s\sqrt{ 4+\tau_s^2}  }{1+\tau_s^2}
\end{equation}
We can now plug in the definition of $\tau_s$, which depends in a non-trivial way on $v_{rel}$, and solve numerically for $v_{rel}$, which closes the whole system of equations, letting us find $Re$, $F_D$, and $\tau_s$. Figure \ref{fig:taus} shows the resulting dimensionless stopping time. Small particles close to the center of the disk have $\tau_s \ll 1$, meaning they are strongly coupled to the gas, with velocities closer to the gas velocity than the Keplerian velocity. Large rocks far from the star are weakly coupled, and move at very nearly the Keplerian velocity.

Since we know the relative velocities, we can evaluate the drag regime in which these particles live by checking the conditions in equation \ref{eq:drag}, and their inward velocity $v_r$ with equation \ref{eq:vr}. We can therefore estimate the timescale on which the particles will lose a substantial fraction of their orbital radius as $t_{in} = r/v_r$, shown in figure \ref{fig:tinflow}. Particles of order 1 meter in size, spiral in to the star in of order 100 years from 1 A.U. This is the classical meter barrier problem.

\section{The Magnus Force}

The lift force experienced by spinning objects has been studied analytically, numerically and experimentally over the past century. A typical case involves a sphere or cylinder spinning about its axis of symmetry suspended in a wind tunnel, with the background velocity perpendicular to the spin axis. The resultant forces on the cylinder can be measured (e.g. by measuring the tension in the apparatus suspending the cylinder) or computed (by integrating the appropriate components of the stress tensor along the surface of the object).

Just as with the drag force, the lift force $F_L$ may then be expressed as a dimensionless coefficient,
\begin{equation}
C_L = \frac{\vec{F}_L \cdot \hat{n}}{(1/2) \pi \rho_g v_{rel}^2 s^2}
\end{equation} 
which should in principle only depend on the dimensionless parameters of the flow, $Re$, $\mathcal{S} \equiv s \omega / v_{rel}$, and $\lambda/s$. Unlike the drag force, which always points towards $-\hat{v}_{rel}$, the lift force refers to any force on the object perpendicular to $\hat{v}_{rel}$, so it is in principle a 2D vector quantity. Though some authors do indeed investigate it this way \citep[e.g.][]{Poon2013}, for simplicity we take $C_L$ to be a scalar quantity with $\hat{n} = \vec{\omega} \times \vec{v}_{rel}/|\vec{\omega} \times \vec{v}_{rel}|$, so that positive (negative) values of $C_L$ refer to lift forces aligned (anti-aligned) with $\vec{\omega} \times \vec{v}_{rel}$. When the spin and velocity axes are aligned, we take $C_L=0$. By symmetry we take the lift force to be entirely parallel or anti-parallel to $\hat{n}$, though clearly this will not always be the case.

Among the best-known results in the long history of the lift force is that of \citet{Rubinow1961}, who analytically computed the flow around a spinning sphere in the limit that $Re\ll 1$. They derived a value for the lift coefficient
\begin{equation}
C_L ^\mathrm{Rubinow}= 2\mathcal{S} | \hat{\omega} \times \hat{v}_{rel} |
\end{equation}
to leading order in $Re$. This is a reasonable approximation to $C_L$ when $Re\la 1$, so long as $\lambda/s < 4/9$. This turns out to be a rare circumstance in our fiducial disk model, in that once $Re\la 1$, typically $\lambda/s \ga 4/9$. When $\lambda/s \ga 4/9$, we assume that $C_L = 0$, since the fluid no longer `cares' about the physical size of the object. This means that $C_L$ will rapidly drop from arbitrarily large values to zero for $Re\sim1$ and large $\mathcal{S}$. The nature of this transition is uncertain and not constrained by data.

At higher $Re$, no analytic results are available. Individual studies will typically run a series of laboratory or numerical experiments over a small dynamic range of $Re$ and $\mathcal{S}$. Many report fitting formulae for their results, though of course experimental error and the choice of functional form can make these unreliable even over the range of parameter space probed by the experiments. For instance, \citet{Oesterle1998} and \citet{You2003} cover a similar range in $Re$ and $\mathcal{S}$, but their formulae disagree by as much as 50\% for $Re\approx 20$ and $\mathcal{S}\approx 1$. 

An extensive review by \citet{Loth2008} has assembled a large quantity of historical data on the lift owing to both particle spin and background shear in the fluid flow. He proposes the following global fit for $C_L$ as a function of $Re$ and $\mathcal{S}$ for low to moderate $Re \la 1000$.
\begin{eqnarray}
C_L^\mathrm{Loth} &=& 2\mathcal{S} [1 - (0.675 + 0.15(\nonumber \\
& & 1+\tanh(0.56(\mathcal{S} - 1)))) \tanh(0.18 Re^{1/2}) ]
\end{eqnarray}
This fit encompasses data from \citet{Legendre1998}, \citet{Bagchi2002}, \citet{Tri1990}, and \citet{Tsuji1985}, and approaches the \citet{Rubinow1961} solution in the limit $Re \rightarrow 0$.

At higher $Re$, the picture becomes murkier owing to the onset of turbulence in the wake of the sphere. For $Re\ga 40,000$, \citet{Loth2008} describes two regimes, both with values of $C_L$ wildly divergent from the formula quoted above when $\mathcal{S} \la 1$. In the subcritical regime, where the surface of the object is smooth enough that the boundary layer remains laminar as it separates from the object, many authors have noted that $C_L$ changes sign \citep{Macoll1928, Davies1949, Tani1950} at low $\mathcal{S}$. \citet{Kim2014} have recently developed a detailed model that predicts where and how this `inverse' Magnus effect occurs based on the results of wind tunnel measurements for $0.6\times 10^5 < Re < 1.8 \times 10^5$ and $0<\mathcal{S}<2$. We expect that this regime is irrelevant for solids in protoplanetary disks owing to the irregular shapes involved, analogous to the dimples on golf balls \citep{Davies1949, Bearman1976}. 

When the particles are rough, or $Re$ is sufficiently large, turbulent rather than laminar separation occurs. In this `super-critical' regime, observational data suggest that $C_L$ can become quite large again, nearly approaching the \citet{Rubinow1961} limit. Fitting to a series of measurements collected using baseballs, \citet{Sawicki2003} suggest
\begin{equation}
C_L^\mathrm{Sawicki} = \mathcal{S} \min(1.5, 0.6+0.09/\mathcal{S} )
\end{equation}
which agrees well with data from \citet{Nathan2006}, \citet{Briggs1959}, and \citet{Watts1987} for $Re > 10^5$. Unfortunately, it is not obvious how one would connect the low $C_L$ at moderate $Re$ to these high values of $C_L$ at high $Re$. Moreover, the \citet{Sawicki2003} fit over-predicts $C_L$ for high $\mathcal{S}$ where the separation between sub- and super-critical regimes is expected to disappear. For the sake of continuity with the moderate-$Re$ regime, we therefore adopt 
\begin{equation}
C_L^\mathrm{Tanaka} = \mathcal{S} \min(0.5, 0.5/\mathcal{S})
\end{equation}
from \citet{Tanaka1990} in the high-$Re$ regime. This is a conservative choice, and it is possible that in this regime we may be underestimating the Magnus force by about a factor of two.

We thereby arrive at the following formula for the lift coefficient, which we expect to be a reasonable approximation for objects in protoplanetary disks over at least 5 orders of magnitude in $Re$ and $\mathcal{S} \la 10$,
\begin{equation}
\label{eq:CL}
C_L =
\begin{cases}
 |\hat{\omega} \times \hat{v}_{rel}| \max( C_L^\mathrm{Loth}, C_L^\mathrm{Tanaka} ) & \mbox{for}\ \lambda/s<4/9 \\
 0 & \mbox{for}\ \lambda/s \ge 4/9
 \end{cases}
\end{equation}
with the caveat that the model of \citet{Kim2014} and the fit from \citet{Sawicki2003} may be relevant for $\mathcal{S} \la 1$ and $Re \ga 10^5$.

The ratio of the lift force to the drag force, assuming that the lift is perfectly aligned with $\hat{n}$, is simply $C_L/C_D$. This quantity should give us an idea of when and where the lift force is important. The three limiting cases (neglecting the orientation factor) are
\begin{equation}
\label{eq:liftRatio}
\frac{C_L}{C_D} \rightarrow
\begin{cases}
\frac1{12} \mathcal{S}\ Re & Re \ll 1\ \mathrm{and}\ \lambda/s<4/9 \\
1.1\ \mathcal{S}  &\mathcal{S} \ll 1\ \mathrm{and}\ Re \gg 1 \\
1.1 &\mathcal{S} \gg 1\ \mathrm{and}\ Re \gg 1 
\end{cases}
\end{equation}
Excluding the possibility of extremely large dimensionless spins, the maximum values of $C_L/C_D$ are likely to be found in the high-$Re$ regime, which tends to correspond to the inner regions of the disk and to larger grains.

\section{The spin rate}
\label{sec:analytic}
\subsection{Empirical values of $\omega$}

We now turn to the question of what value to use for $\omega$.  There is a substantial quantity of data available \citep{warner2009}, derived largely from photometric light curves, regarding the spin states of modern asteroids with sizes $s \ga 10$ meters, though of course these asteroids have been subject to several Gyr of evolution under a different set of environmental conditions than existed in the early solar system. These data show (see Figure \ref{fig:sizeOmega}) that asteroids larger than 100 m have a minimum period of a few hours, consistent with the idea that they are largely held together gravitationally, in which case the minimum period is roughly $\sqrt{3\pi/\rho_s G}$, where $\rho_s$ is the mean density of the body. Below 100 meters, there are plentiful rapidly-rotating asteroids, including some with periods of a minute or less. With this in mind we adopt the following functional form
\begin{equation}
\label{eq:omegaEmpirical}
\omega_{\mathrm{empirical}} = \begin{cases} \omega_0 \left(\frac{s}{s_\mathrm{rubble}}\right)^{-\beta} & s<s_\mathrm{rubble} \\
\omega_0  & s>s_\mathrm{rubble} \end{cases}
\end{equation}
Above a critical size $s_\mathrm{rubble}$ we take particles to be rubble piles held together by gravity with a constant rotation period $\omega_0$ of order the free fall time. Smaller particles are taken to have a power law index $-\beta$ with smaller grains rotating faster. 

The black and red dashed lines in figure \ref{fig:sizeOmega} show examples with $(\omega_0, s_\mathrm{rubble}, \beta) = (3\times 10^{-4}$ seconds$^{-1}, 1\ \mathrm{km}, 1.0)$, the shallow case, and $(3\times 10^{-4}$ seconds$^{-1}, 0.3\ \mathrm{km}, 1.5)$, the steep case, respectively. These lines are plausible, though simplified, representations of the data. We are mostly concerned with much smaller-sized objects, so even if we believed the size-frequency distribution of solids in the solar system to be the same today as it was in the presence of a gaseous disk, these relations would still be huge extrapolations owing to the difficulty of observing small solar system bodies. There is other evidence that small meteor-sized objects may rotate rapidly, namely the appreciable fraction of non-linear meteor trails \citep{beech1988}. 

\begin{figure}
\centering
\includegraphics[width=9cm,clip=true,trim=0cm 0cm 0cm 0cm]{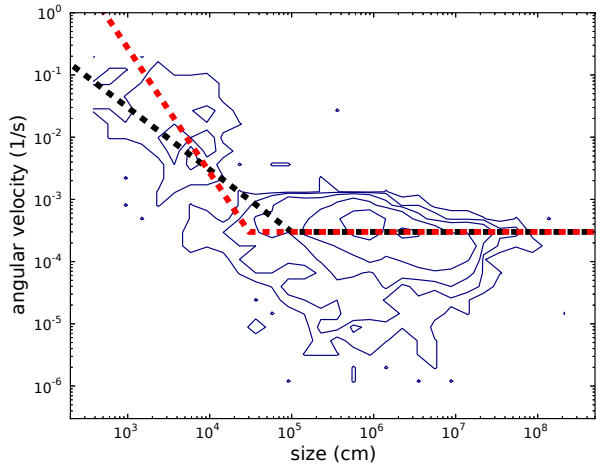}
\caption{The observed angular velocity-size relation of modern asteroids from \citet[][February 2014 version]{warner2009}. The logarithmically-spaced contours show the density of the highest-quality ($U>2$) data points in the catalog where both parameters have measured values, rather than just upper or lower limits. The black and red dashed lines show equation \ref{eq:omegaEmpirical} evaluated with the same $\omega_0=3\times 10^{-4} $ seconds$^{-1}$, but with two different power law slopes ($-1$ and $-1.5$) and two different cutoffs (1 km, 0.3 km).}
\label{fig:sizeOmega}
\end{figure}

It is worth pointing out that if $\beta>1$, eventually the grains will reach their theoretical maximum spin rate of $\omega s \sim 1 $ km s$^{-1}$, the sound speed of solid rock. This will happen for $s \la s_\mathrm{rubble} (\omega_0 s_\mathrm{rubble} / 1\ \mathrm{km}\ \mathrm{s}^{-1})^{1/(\beta-1)}$. In subsequent figures, we cap $\omega s$ at 0.01 km s$^{-1}$, which should be substantially slower than the maximum spin rate, though still perhaps implausibly large.

For these values of $\omega$ as a function of the object's size, we can estimate the ratio of the Magnus force to the drag force as a function of quantities we know how to calculate in the classical scenario.

\begin{figure}
\centering
\hspace*{-1.05cm}\includegraphics[width=8cm,clip=true,trim=1.5cm .3cm .8cm .3cm]{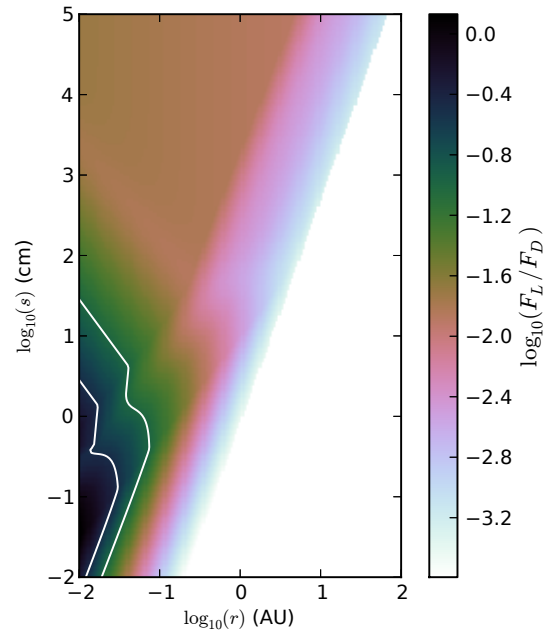}
\caption{The log of the ratio of the Magnus force to the drag force, with contours at $-1$ and $-0.5$. The spin rate decreases as $s^{-1}$, consistent with data for modern asteroids. The Magnus force under this scenario is not particularly effective for meter-sized objects, but may be promising for small grains close to the star. The white region at large radii corresponds to $Re<1$, where we believe the Magnus force is unlikely to play a role.}
\label{fig:fratio}
\end{figure}

\begin{figure}
\centering
\hspace*{-1.05cm}\includegraphics[width=8cm,clip=true,trim=1.5cm .3cm .8cm .3cm]{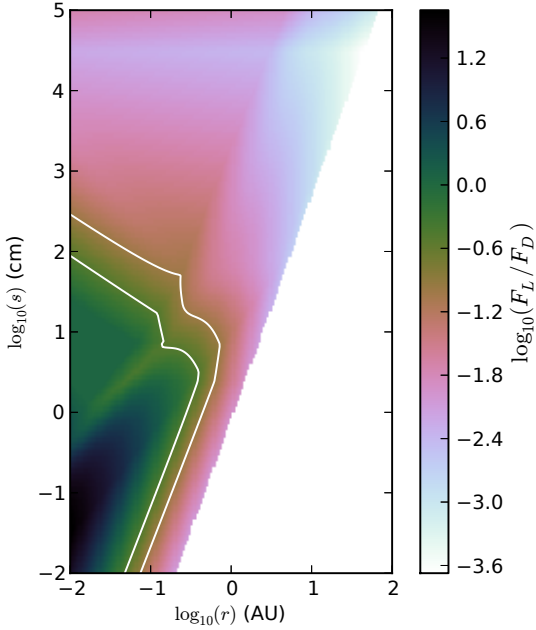}
\caption{The log of the ratio of the Magnus force to the drag force, with contours at $-1$ and $-0.5$. The spin rate decreases as $s^{-1.5}$, again consistent with data for modern asteroids. This case shows similar behavior to the shallower $\omega(s)$ distribution, though here the steeply increasing spin rate for smaller grains means that grains smaller than 10 cm are strongly affected by the Magnus force, provided they have a high enough $Re$ for such fluid effects to be important.}
\label{fig:fratioSteep}
\end{figure}

Clearly this ratio depends quite sensitively on what we assume about the size-frequency distribution -- the Magnus force may be small but not negligible in one case, or overwhelmingly large in the other. Figure \ref{fig:fratio} shows the ratio for the shallow $\omega(s)$ case, while \ref{fig:fratioSteep} shows the steeper case. When $\omega \propto s^{-1}$, the features in this graph come mostly from $v_{rel}$ -- when the relative velocity of the particle and the gas is minimized, $F_L/F_D$ is maximized. This effect can be completely swamped by any change in the $\omega(s)$ distribution, as in figure \ref{fig:fratioSteep}, where rapidly rotating well-coupled ($\tau_s \ll 1$) grains show the most prominent influence of the Magnus force.

\subsection{Theoretical estimates for $\omega$}

Given the uncertainty in using the spin rate distribution $\omega_\mathrm{empirical}$, it is also worth considering the physical mechanisms responsible for the spin. This has its own set of uncertainties, including unknowns in the basic physics, and the sensitivity some of these physical mechanisms have to unknown quantities, such as the small-scale shapes of the solid bodies, or the instantaneous distribution of particle spins, sizes, and orbits. Nonetheless we discuss a few potentially important ingredients. 

\subsubsection{YORP}
A great deal of effort has gone into explaining the spin of modern asteroids with the YORP effect \citep{rubincam2000,vokrouhlicky2002}, caused by the anisotropic re-radiation of incident sunlight. Theoretically this torque scales roughly as
\begin{equation}
\mathcal{T}_Y \approx \frac{B G_1 }{4 a^2\sqrt{1-e^2}\rho_{s} s^2}C_Y I
\end{equation}
following \citep{rossi2009}. The orbital semi-major axis and eccentricity, $a$ and $e$ appear in the denominator, making the force more effective near the center of the disk for particles on eccentric orbits. The solar radiation constant $G_1$ is taken to be $10^{22}$ g cm s$^{-2}$, while $B\approx2/3$ is the Lambertian scattering coefficient. The coefficient $C_Y$ is a number variously measured for solar system asteroids to be between -0.025 and 0.025 \citep{rozitis2013}, while the moment of inertia $I \approx (\pi/2) \rho_{s} s^5$. 

We therefore expect the timescale on which a particle's angular velocity may be changed by YORP is
\begin{equation}
\label{eq:tYORP}
t_\mathrm{YORP} \sim \omega I/|\mathcal{T}_Y| \sim 10^{6} \omega_0 r_{AU}^2 \rho_{s,0} s_0^2 \ \mathrm{seconds}
\end{equation}
where we have adopted $C_Y = 0.01$, and normalized $\omega$, $\rho_s$ and $s$ to unity in cgs. Modern asteroids exhibit $C_Y$ of both signs, meaning YORP can spin up an asteroid (which if not countered by some other torque may lead to its disruption \citep{jewitt2014,jacobson2014}) or spin it down to near-zero $\omega$, perhaps explaining the population of extreme slow rotators.

YORP has the disadvantage of requiring direct irradiation by the star, which may be impossible for much of the disk. A gaseous disk with homogeneously-mixed $0.1 \mu m$ dust can be quite optically thick ($\tau \sim 10^5)$ even in the vertical direction at optical wavelengths \citep{Chiang1997}. In order for YORP to be effective, these small grains would need to be depleted in favor of much larger grains, since the opacity in the geometric limit scales as the surface area to volume ratio of an individual grain.

\subsubsection{Hydrodynamic friction}

A torque unlikely to be important in today's solar system due to the absence of gas, but likely quite important during the disk's lifetime, arises from drag along the surface of the body. It is not precisely clear how to estimate this torque, as it may well depend on the shape and small-scale features of the object, not to mention the detailed flow of the gas. As a rough estimate, we take
\begin{eqnarray}
\chi &=& | \hat{\omega} \times \hat{v}_{rel} | \\
F_{top} &=& \epsilon F_D(v = |v_{rel} \chi + \omega s |) \\
F_{bot} &=& \epsilon F_D(v= |v_{rel} \chi - \omega s|) \ \mathrm{sign}(\omega s - v_{rel}\chi )\\
\mathcal{T}_D &=& -s | F_{top} + F_{bot} | /2
\label{eq:dragTorque}
\end{eqnarray}
Here $\chi$ is just the fraction of the velocity that is not aligned with the spin axis, and should therefore be included in the relative velocity between the object and the surrounding gas. $F_{top}$ and $F_{bot}$ are the tangential forces exerted on the skin of the object on the side rotating in to the oncoming relative velocity and the side heading away from that velocity respectively. These are estimated to be some fraction $\epsilon$ of the overall drag force the object as a whole would feel if it were moving at the surface velocity relative to the gas. The sign change when $s\omega = v_{rel}$ comes about because in the limit of fast rotation, both forces will oppose that rotation, whereas for slow rotation the forces should nearly cancel each other out, until at zero rotation this drag torque is zero.

In the limit where the particle is spinning rapidly, i.e. $\mathcal{S} \gg 1$, we find that $\mathcal{T}_D = -\epsilon C_D  \rho_g \pi s^5 \omega^2  $, where of course $C_D$ depends on the velocity $\omega s$ and location in the disk via the Reynolds number. We can compare this to an expression from \citet{Loth2008} for the torque in the limit where $v_{rel} \rightarrow 0$ and the background shear is zero,
\begin{equation}
\mathcal{T}_{D,Loth} = -\pi 16 \rho s^5 \omega^2 Re_\omega^{-1}\left(1 + \frac{5}{64\pi} Re_\omega^{0.6}\right).
\end{equation} 
We use $Re_\omega$ to denote the Reynolds number where the characteristic velocity is $\omega s$ instead of $v_{rel}$. This agrees with the expression for $\mathcal{T}$ given by \citet{Rubinow1961} for $Re_\omega \ll 1$. In this regime, we know that $C_D = 24/Re$, so the two expressions agree for $\epsilon=2/3$. This gives us some confidence about our expression for $\mathcal{T}_D$, although it still remains an extrapolation to apply it when $\mathcal{S} \la 1$.

In the high-spin regime, the spindown time is 
\begin{equation}
\label{eq:tSpindownFast}
t_\mathrm{spindown, fast} = \omega I/|\mathcal{T}_D| \sim \omega^{-1} \frac{\rho_s}{\rho_g} \frac{1}{2 \epsilon C_D}  
\end{equation}
In other words the particle will spin for some large multiple of its rotational period, of order or a bit larger than the density contrast. In the opposite limit $\mathcal{S}\ll 1$, the drag torque is a little more complicated according to equation \ref{eq:dragTorque},
\begin{equation}
\mathcal{T}_{D,\mathrm{slow}} \sim  \epsilon C_D \pi \chi^2 \rho_g s^4 v_{rel} \omega \left(1 + \frac12 \frac{\partial \ln C_D}{\partial\ln v} |_{v=v_{rel}} \right) 
\end{equation}
Note that in general the logarithmic derivative will be between $-1$ and $0$ depending on the drag regime. For convenience we will abbreviate $\zeta = (1+(1/2) \partial \ln C_D/\partial \ln v |_{v=v_{rel}})$, which will always be of order unity. In this limit the spindown time is
\begin{equation}
\label{eq:tSpindownSlow}
t_\mathrm{spindown,slow}= t_\mathrm{spindown,fast} \chi^{-2} \mathcal{S}  \zeta^{-1}
\end{equation}
i.e. substantially shorter than in the fast-spinning limit. This is somewhat counter-intuitive because in this limit the two drag torques are acting in opposite rotational directions instead of acting together to slow down the rotation. However, the drag in this regime is much more effective (all other parameters being equal), because the drag forces $F_{top}$ and $F_{bot}$ care about $v_{rel}$, which is large compared to the velocity which is being decreased, $\omega s$.

Comparing equation \ref{eq:tYORP} and equations \ref{eq:tSpindownFast} and \ref{eq:tSpindownSlow} we see that each timescale has a different dependence on $\omega$, with the YORP timescale increasing with increasing $\omega$, but the spin down times decreasing or remaining constant with increasing $\omega$. This means that if these two forces were to counteract each other, they would pick out an equilibrium $\omega$, which we can find by setting the timescales equal. It turns out that in this equilibrium, the slow-rotation limit is appropriate, in which case we obtain
\begin{equation}
 \omega_0  =\frac{r_{AU}^{13/7}}{6.65} \frac{(\tau_s^{-1} + \tau_s) / \sqrt{4+\tau_s^2}}{2 \epsilon C_D s_0 \chi^2 \zeta}
\end{equation}
Remarkably, if we focus on the high-$Re$ and loosely coupled ($\tau_s \gg 1$) regime, and evaluate for $a = 3$ AU, appropriate for the modern asteroid belt, we obtain $\omega_0 \sim 2.6 / \epsilon \chi^2 s_0$, numerically very similar to the shallow $\omega_\mathrm{empirical}$ case when $\chi\sim\epsilon\sim 1$. 

We have so far assumed that surface drag will monotonically spin down a particle, but this is not guaranteed. For particles with asymmetries derived from an empirical model, \citet{apek2014} has found that appreciable spin can be obtained by the action of hydrodynamic drag in the context of cometary meteoroids. The median spin rate for particles spun up in this manner is
\begin{equation}
\mathcal{S} \sim  10^{-3} s_0^{0.12}.
\end{equation}
The deviation from $\mathcal{S}\sim const.$ arises from the assumed scaling of asymmetries in particle shape as a function of size. The simulations are also carried out in a low-$Re$ regime where we expect the Magnus force to be unimportant, Nonetheless, this may provide an approximate lower limit on the spin rate. Moreover, although the median is rather small, the distribution is quite broad -- log-normal with 0.5 dex scatter. Particles in the simulations in question can even reach breakup spin rates.

\subsubsection{Collisions}

The gravitational torques exerted by the star (or planets) on planetesimals, commonly included in evolutionary models for the asteroid belt, are probably unimportant given the long timescales over which they act and the smallness of the objects in which we are most interested. Similarly, accretion torques and shear of the disk across the size of the solid body should be small effects. Collisions, however, may be important or even dominant in determining the spin rate distribution, but modeling them may require a self-consistent evolution of a spatially-dependent size-frequency distribution. Nonetheless we can take some straightforward steps in that direction.

We begin by adopting some idealizing assumptions which can be relaxed later. We assume that all particles are the same size with an isotropic velocity dispersion $\sigma$, local number density $n$, and cross-sectional area $\pi s^2$. In this case the characteristic time between collisions for a particular particle is $t_c \sim (n \sigma \pi s^2)^{-1}$. If the particles are in a thin disk, we can take the velocity dispersion to be $\sigma = v_K H/r = \Omega_K H$, where $H$ is the scale height of the particles. The column density of particles (number of particles per unit area) is just $N_s = n H$, so $t_c \sim (N_s \pi s^2 \Omega_K)^{-1}$. By assuming all particles to be the same size, we can estimate the column density of particles to be the mass surface density of solids divided by the mass of an individual particle, i.e. $N_s = 3\Sigma_s/(4\pi \rho_s s^3)$. Adopting the mass surface density of solids 
\begin{equation}
\Sigma_s = 33 F Z_{rel} r_{AU}^{-3/2}  \mathrm{ g\ cm}^{-2} 
\end{equation}
from \citet{chiang2010}, where $Z_{rel}$ is the metallicity relative to solar, we find the following typical time between collisions,
\begin{equation}
\label{eq:tColl}
t_c  \sim 0.04 (F Z_{rel})^{-1} r_{AU}^{3/2} s_0 \rho_{s,0} / \Omega_K.
\end{equation}
In other words, particles can collide more frequently than an orbital time for small particles near the center of the disk.

Perhaps more important than the orbital time are the other timescales on which we expect the spin may change. We can envision a scenario in which collisions at some velocity $v_c$ set a maximum spin rate of $\omega_c \sim v_c/s$, which then decays on a spindown timescale. Particles can reach this maximum spin rate following a collision at a large impact parameter with an object of similar size, so long as a substantial fraction of the impactor's angular momentum in the target's reference frame ends up as spin angular momentum of the target \citep{farinella1992}. For smaller fragments of a collision, rapid rotation is also a common outcome \citep[e.g.][]{Fujiwara1981, Paolicchi1989}. 

The typical spin rate for objects subject to such collisions will then either be $\omega_{c}$ if $t_c \ll t_{spindown}$, or a much smaller value set by some other process (e.g. the YORP-spindown balance posited above) when $t_c \gg t_{spindown}$. The ratio of these two timescales evaluated at $\omega = \omega_c$ is roughly
\begin{equation}
\frac{t_c}{t_\mathrm{spindown}} \sim   \frac{\epsilon C_D}{ Z_{rel}} v_{c,3}  r_{AU}^{3/14} 
\end{equation}
where we have normalized the collision velocity to $10^3$ cm/s. Enhanced particle densities, as commonly present in direct gravitational collapse or streaming instability scenarios, will decrease this ratio in direct proportion to the overdensity. We therefore consider it quite plausible that collisions may become frequent enough over the course of the planet formation process that even effective gas drag would be insufficient to slow the particles much below their maximum spin value $\omega_c$. If we make the further approximation that $v_c \sim v_{rel}$, the dimensionless spin rate would approach $\mathcal{S}\sim 1$. In this scenario, equation \ref{eq:liftRatio} tells us that the ratio of the lift to the Drag force would be of order unity for grains larger than the mean free path of gas particles in the disk.

\section{Numerical Integrations}
\label{sec:numerical}

\begin{figure*}
\centering
\vspace*{-0.1cm}\includegraphics[width=16cm]{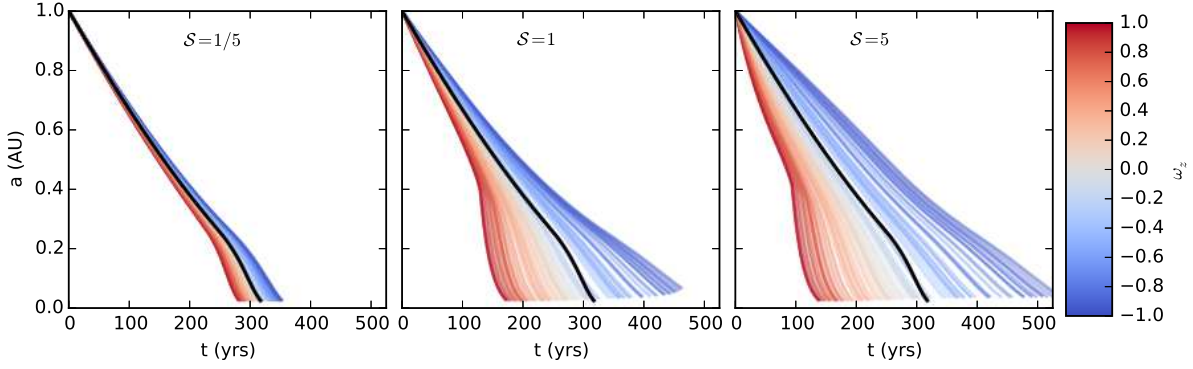}
\caption{The semi-major axis as a function of time for 100 particles with randomly-oriented spins. Each panel shows a different value for the initial spin rate, while the color indicates how well the spin-axis is aligned with the orbital angular momentum. The black line shows the case with zero spin for reference. Successively faster and more anti-aligned spins both tend to increase the lifetime of a particle in the disk. For very rapid spins, the orientation starts to matter much less, and all particles survive much longer than in the reference case.}
\label{fig:a}
\end{figure*}

\begin{figure}
\centering
\hspace*{-.01cm}\includegraphics[width=9cm]{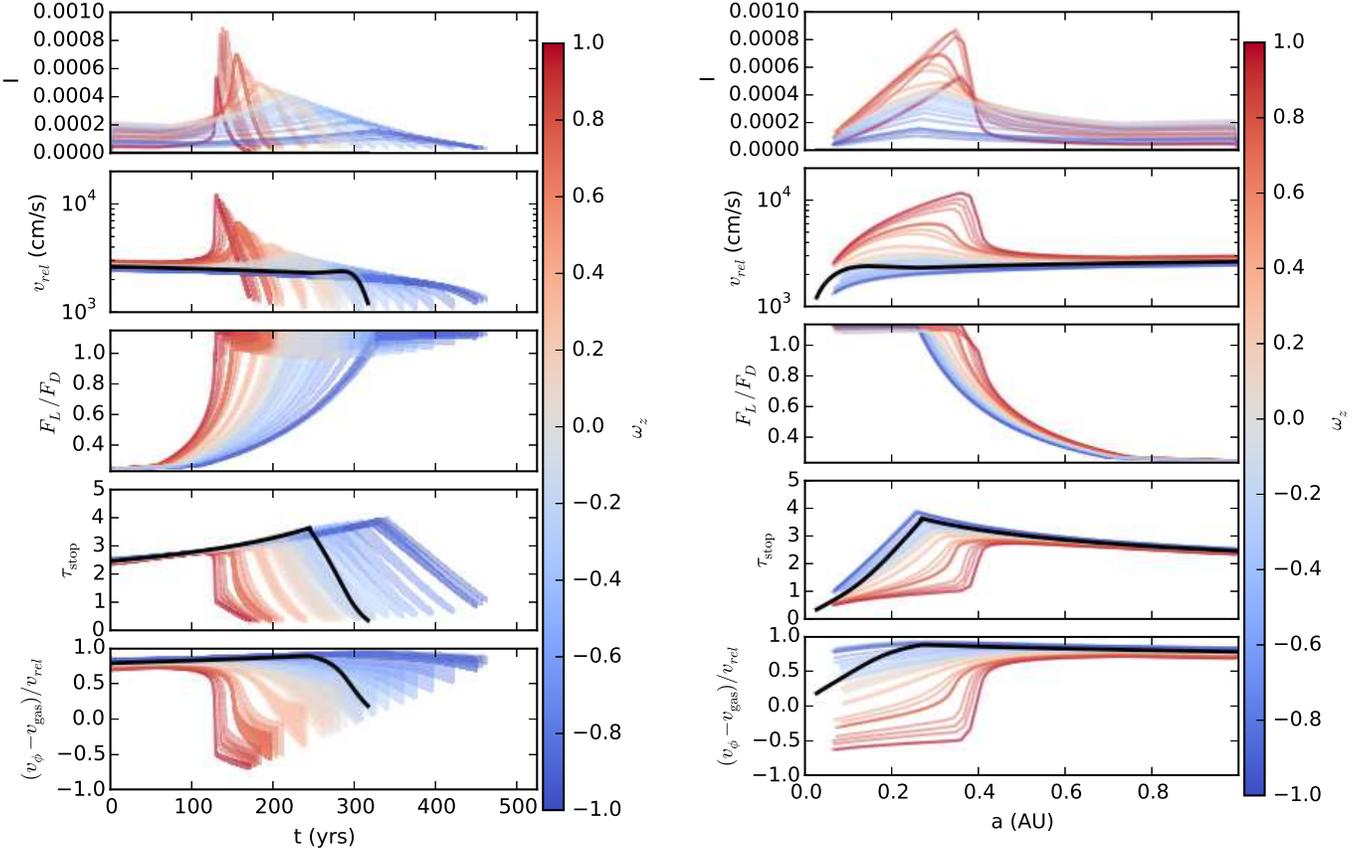}
\caption{Various properties of the integrations for $\mathcal{S}_\mathrm{target}=1$. As in Figure \ref{fig:a}, the colors represent the alignment of the spin with the orbital plane, and the no-spin case is shown as the black line. From top to bottom, the quantities are inclination, relative velocity, Magnus force ratio, dimensionless stopping time, and the normalized azimuthal component of $v_{rel}$. The Magnus force induces oscillations in all of these quantities, and substantial departures from the no-spin case.}
\label{fig:r1_raw}
\end{figure}

\begin{figure}
\centering
\hspace*{-.01cm}\includegraphics[width=9cm]{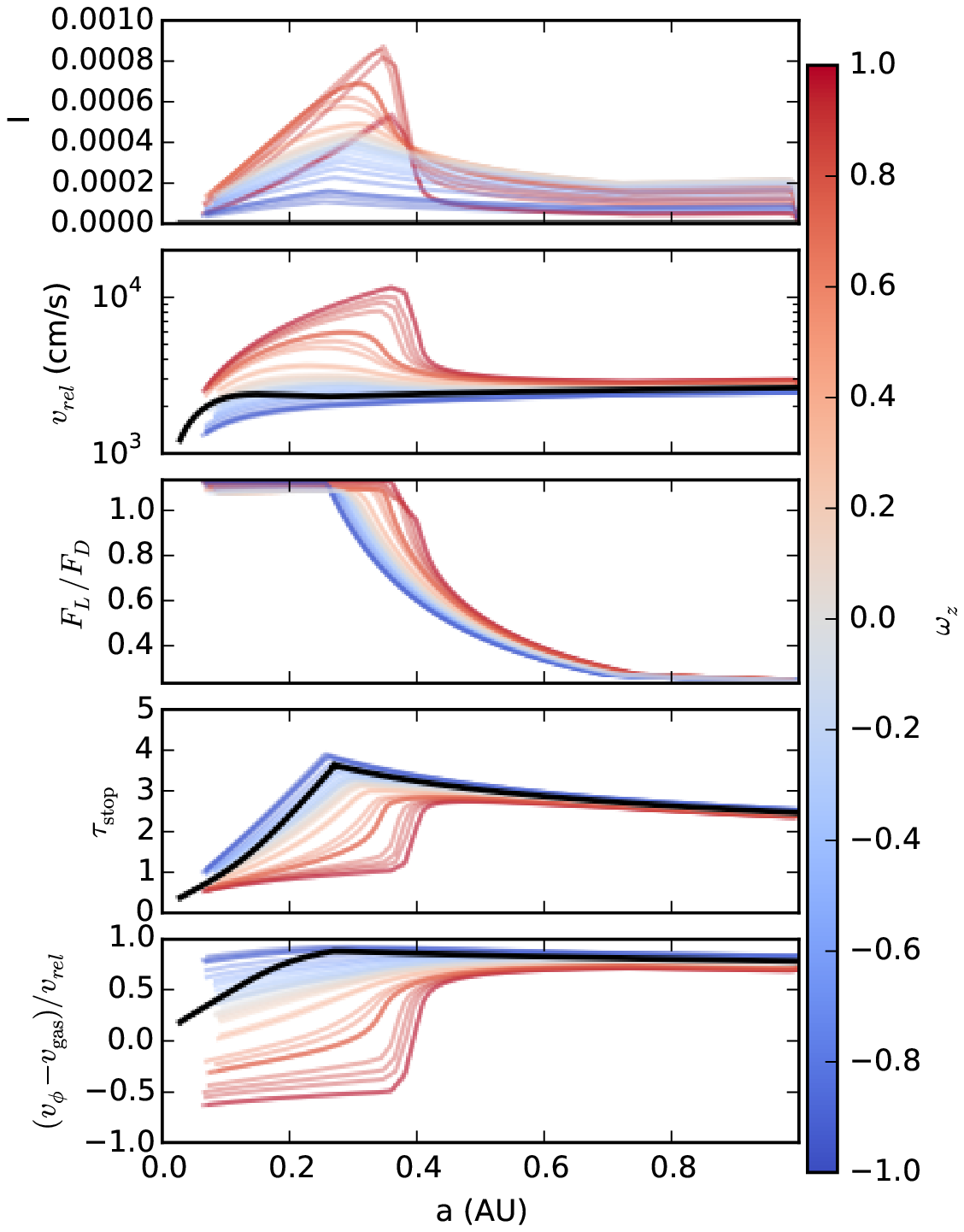}
\caption{Various properties of the integrations for $\mathcal{S}_\mathrm{target}=1$. Same as Figure \ref{fig:r1_raw}, except that the quantities have been time-averaged in 1-year windows and each quantity is shown as a function of semi-major axis instead of time. Particles therefore traverse this plot from right to left. This smooths out the oscillations seen in the previous figure, and accounts for the fact that at a given time the particles will be at different locations in the disk, and hence subject to different densities, gas velocities, etc. The solid black line again shows the zero-spin trajectory. }
\label{fig:r1_avg}
\end{figure}

We have shown that solid particles in our fiducial protoplanetary disk can plausibly have spins large enough for the lift force and the drag force to be comparable. The next step is to understand the effect this new force has on the dynamics of single grains. To that end, we carry out explicit numerical integrations of the equations of motion for individual particles traveling through the simple laminar disk model presented in section \ref{sec:classical}. 

In section \ref{sec:analytic}, we saw that there are many plausible ways to assign a spin $\omega$ to particles in the disk, depending on their size, location, and most importantly the physical processes that are relevant in determining the spin. In order to carry out integrations, we need to make an explicit choice. The simplest approach would be to pick a constant value of $\omega$ at the beginning of the simulation and leave it unchanged throughout. However, since several scenarios pick out a particular dimensionless spin $\mathcal{S}$, rather than a dimensional spin $\omega$, we set the spin rate such that $\mathcal{S}$ will exponentially approach some constant $\mathcal{S}_\mathrm{target}$, and we run several sets of simulations with different values of $\mathcal{S}_\mathrm{target}$.

We also expect that the orientation of the spin axis $\hat{\omega}$ will evolve over time. Each time a particle collides, or is subject to enough time under hydrodynamical or radiative torques, the spin may be reoriented. However, we have found that keeping the spin axis constant in time makes interpreting the results of the simulations more straightforward. This is because, as we shall see, the particle's behavior is largely determined by $\hat{\omega}_z$, the projection of the spin orientation on the angular momentum vector of the gaseous disk. One should therefore keep in mind that we do not expect any particular particle to follow these exact trajectories. Instead, each particle behaves according to its instantaneous semi-major axis and spin orientation.

We evolve each particle's position, velocity, and spin in an inertial frame with Cartesian coordinates $x$, $y$, and $z$. The disk is centered at the origin with its angular momentum parallel to $\hat{z}$. In this system, each particle is subject to the following 7 coupled ODE's
\begin{eqnarray}
\frac{d\vec{x}}{dt} &=& \vec{v} \\
\frac{d\vec{v}}{dt} &=& -GM_* \frac{\vec{x} }{||\vec{x}||^3} + \frac{\vec{F}_D}{m_s}  + \frac{\vec{F}_L}{m_s} \\
\frac{d |\vec{\omega}|}{dt} &=& (\mathcal{S}_\mathrm{target} - \mathcal{S}) \frac{|\vec{\omega}|}{t_\mathrm{orb}}
\end{eqnarray}
Here the particle mass $m_s=(4\pi/3)\rho_s s^3$ with $\rho_s=1\ \mathrm{g\ cm}^{-3}$. The particle's spin is set to approach a target dimensionless spin $\mathcal{S}_\mathrm{target}$ on a local orbital timescale $t_\mathrm{orb} = \Omega_K^{-1}$. For the sake of simplicity and clarity, the orientation of the particle's spin and the particle's size are taken to be constant throughout the simulation.

The drag and Magnus forces, $\vec{F}_D$ and $\vec{F}_L$ are calculated through equations \ref{eq:drag} and \ref{eq:CL}. Each depends on $v_{rel}$, which is calculated instantaneously as the vector difference between $\vec{v}$ and $\vec{v}_g$, where $\vec{v}_g = (1-\eta) v_K \hat{\phi}$. We initialize the integrations at the mid plane with $y=z=v_z=0$, and $x = 1$ A.U. The in-plane velocities are initialized according to equations \ref{eq:vr} and \ref{eq:vphiRel}, i.e. $v_y = v_r$ and $v_x=v_\phi$. The equations are integrated for each particle until it reaches 0.1 AU in semi-major axis, or until the integrator has taken four million steps.

We carry out several hundred integrations\footnote{We employ the publicly-available bsint package available from https://github.com/alrexrudy/bsint. We have verified that it preserves orbital parameters to $\sim10^{-11}$ over $3000$ orbits for our setup without the drag and Magnus forces, i.e. the two-body problem.} with $s=1\ \mathrm{meter}$, since in the fiducial disk model these are the particles that traverse the $\tau_s=1$ line (see figure \ref{fig:taus}) from the weakly- to the strongly-coupled regime. We carry out three different sets of integrations with different initial values of $\mathcal{S}_\mathrm{target}$, shown in the three panels of Figure \ref{fig:a}. The orientation of the spin axis $\hat{\omega}$ is chosen from the uniform distribution over the surface of the unit sphere. In Figure \ref{fig:a}, the color shows the z-component of this orientation, $\hat{\omega}_z$, so that red lines show spins aligned with the orbital angular momentum vector, while blue lines show anti-aligned runs, and lightly-colored lines have a spin-axis nearly lying in the orbital plane.

In all cases the spin has a noticeable effect on the particles' trajectories, conceivably extending the particle's lifetime in the disk by factors of 2 if the spin orientation remained constant. We see that the inward drift velocity of the particles is basically determined by $\omega_z$, with spin-aligned particles flowing in fastest, and progressively more anti-aligned particles flowing in more slowly. This is an intuitive result, since the Magnus force for aligned particles will point towards the star, increasing the circular velocity it needs to stay in orbit at a given semi-major axis. This increase in the velocity will increase the drag force, and shorten the inflow time.

Figure \ref{fig:r1_raw} shows more details of the $\mathcal{S}=1$ run. From top to bottom, the panels are inclination, velocity relative to the gas, the ratio of the lift to the drag force, the dimensionless stopping time, and the fraction of the relative velocity in the azimuthal direction. In this more detailed view of the integrations, $\hat{\omega}_z$ once again determines the trajectory of the particle. For many of the quantities, this is simply a result of the aligned (red) particles reaching the center of the disk more quickly, with the anti-aligned (blue) particles behaving quite similarly later on. To account for this common structure of the trajectories, we can plot each quantity as a function of semi-major axis instead of time, averaging over some time period to remove the oscillations visible in Figure \ref{fig:r1_raw}.

The result is shown in Figure \ref{fig:r1_avg}, where each quantity has been averaged over a moving 1-year window. Qualitatively the paths of the particles do get closer together. However, many of the quantities still display substantial differences as a result of the particles' different spin orientations. At a fixed radius, the aligned particles have higher inclinations, velocities relative to the gas, and $F_L/F_D$. 

Perhaps most dramatically, the aligned particles experience negative values of $v_\phi - v_{gas}$. In other words, the particle experiences a tailwind, yet still spirals in towards the central star. In fact, these particles spiral into the star more quickly than an analogous non-spinning particle. We can also clearly see a qualitative change in the behavior of the particles around $a=0.4$, particularly the spin-aligned particles. At this point, meter-sized particles are crossing in to the high-$Re$ regime, wherein $F_L/F_D$ approaches its maximum value (see equation \ref{eq:liftRatio}). We will investigate this further analytically in the next section.

\section{Implications for Planetesimal Formation}
\label{sec:discussion}

\begin{figure*}
\centering
\hspace*{-.01cm}\includegraphics[width=15cm]{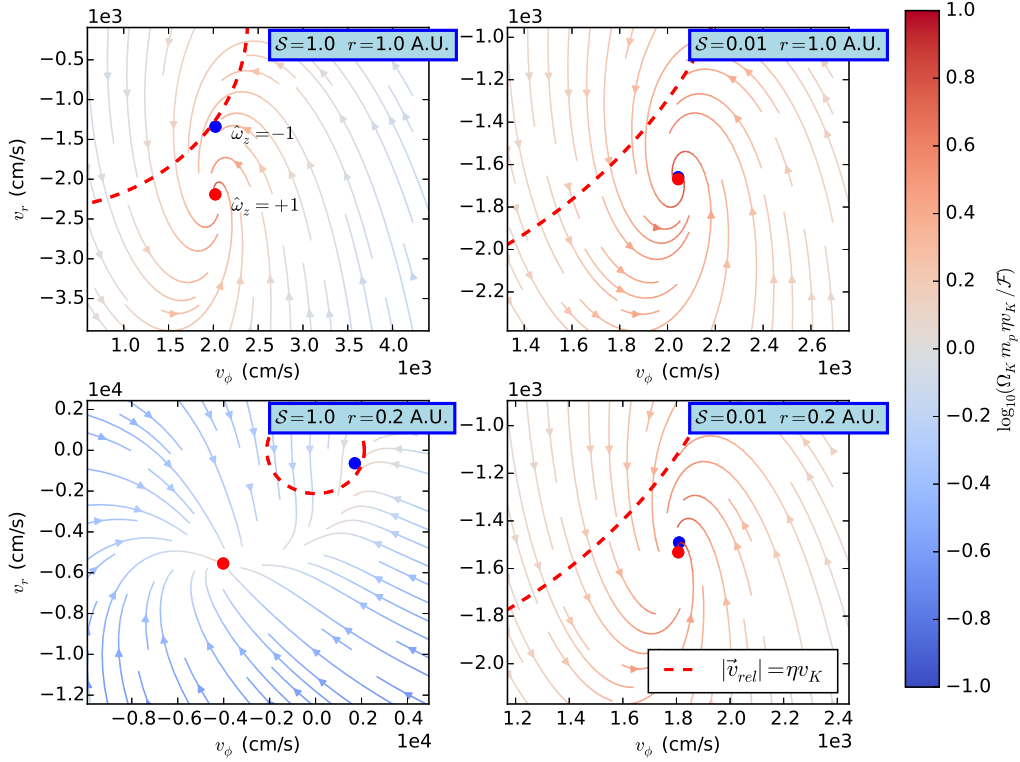}
\caption{The vector field $\mathcal{F}$ as a function of $v_{rel}$ in the frame co-rotating with the gas. $\mathcal{F}$ is evaluated taking $\hat{\omega}_z=+1$, $s=1$ meter, $\rho_\mathrm{grain}=1\ {\rm g}\ {\rm cm}^{-3}$, and $M_*=1 M_\odot$. The left column shows a scenario with strong spin, $\mathcal{S}=1$, while the right column shows $\mathcal{S}=0.01$. Each row shows a different heliocentric radius. The root of $\mathcal{F}$ is marked with a red symbol, while the root of the analogous vector field evaluated with $\hat{\omega}_z=-1$ is shown in blue. The difference in velocity between these two roots in each panel accounts for most of the velocity dispersion seen in the numerical integrations. We also show for reference a red dashed line representing a circle of radius $\eta v_K$ centered at zero relative velocity. The color of the streamlines denotes the timescale for a particle to move a distance $\eta v_K$ in this space. The timescales are typically a few orbital times or less, meaning that particles can quickly reach these equilibrium points.}
\label{fig:stream}
\end{figure*}

In the previous section we saw a variety of intriguing phenomena in our numerical simulations. Here we attempt to understand these results in more detail and their implications for planet formation scenarios. Examining Figure \ref{fig:r1_avg}, we see that the spread in velocities, like essentially all of the time-averaged quantities, is determined by $\omega_z$. Moreover, the spread in averaged relative velocities increases substantially as the particles make the transition from the weakly-coupled $\tau_s>1$ to the strongly-coupled $\tau_s<1$ regime, but then decreases again at small heliocentric radii. In the following we show that these results can be explained quantitatively by finding the force-free (equilibrium) velocities in a co-rotating reference frame. We then use this result to extend our simulation results to a much wider variety of particle sizes and heliocentric radii.

\subsection{Analysis in the co-rotating frame}

\begin{figure*}
\centering
\hspace*{-.01cm}\includegraphics[width=15cm]{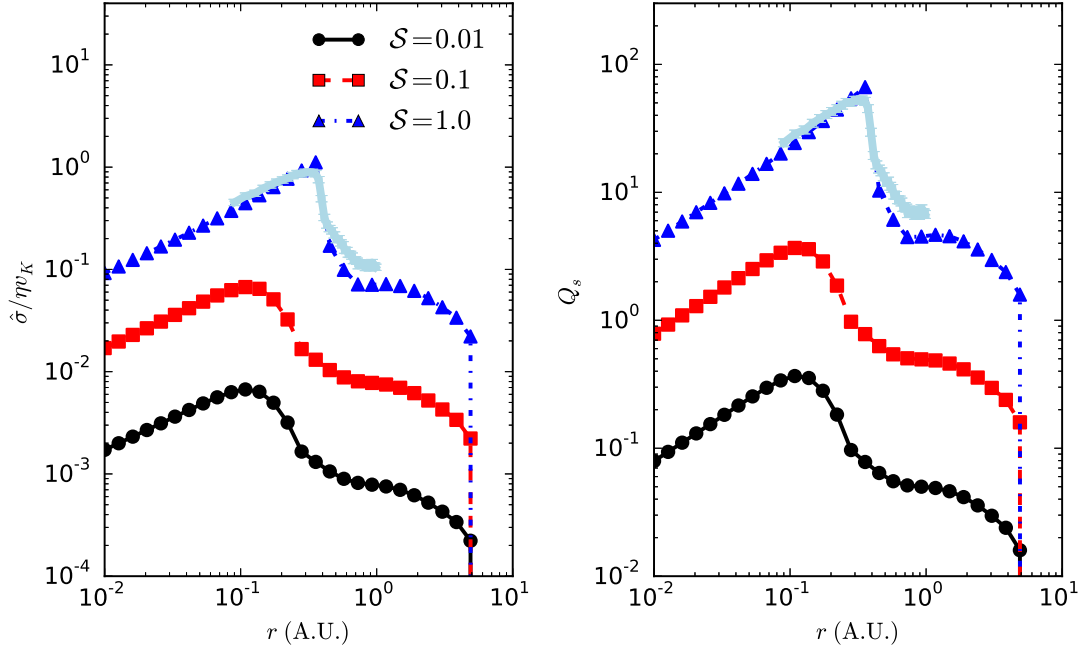}
\caption{The velocity dispersion of particles as a function of radius, normalized to $\eta v_K$, the difference between the gas and Keplerian orbital velocities (left panel), and the resulting value of the Toomre $Q$ parameter for the dust disk (right panel). Lines are shown for different values of the dimensionless spin $\mathcal{S}$. The black, red, and blue lines use Equation \ref{eq:sigmahat} to estimate $\sigma$ in both panels, i.e. the velocity dispersion is assumed to arise purely from the difference between the equilibrium $v_{rel}$ of particles aligned or anti-aligned with the disk angular momentum. The light blue line with error bars shows the velocity dispersion of the ensemble of 1-meter particles whose trajectories were numerically integrated in Section \ref{sec:numerical}. This velocity dispersion is close enough to the approximation of Equation \ref{eq:sigmahat}, that we will adopt $\sigma \approx \hat{\sigma}$ (which is easy to calculate) throughout the rest of this work. }
\label{fig:sigplot}
\end{figure*}

In a rotating frame, the force on a particle can be written
\begin{equation}
\vec{\mathcal{F}} = -\frac{G M_* m_p \vec{x}}{\left(r^2+z^2\right)^{3/2}} + \vec{F}_D + \vec{F}_L  + \vec{\mathcal{F}}_{\rm fict}
\end{equation}
where $\vec{x}$ is the position of the particle, and $\vec{v}$ is its velocity. The cylindrical radius is $r=\sqrt{x^2+y^2}$. The last term represents the ``fictitious'' Coriolis, centrifugal, and Euler forces,
\begin{equation}
\vec{\mathcal{F}}_{\rm fict} = - 2 m_p \vec{\Omega} \times \vec{v} - m_p \vec{\Omega}\times (\vec{\Omega} \times \vec{x}) - m_p \frac{d \vec{\Omega}}{dt} \times \vec{x}
\end{equation}
The drag and Magnus forces are of course dependent on the velocity and other variables. If we narrow our focus to a frame co-rotating with the gas, i.e. $\vec{\Omega} = (1-\eta) \Omega_K \hat{z}$, then $\vec{v}=\vec{v}_{rel}$. This is convenient because both of the hydrodynamic forces depend directly on $v_{rel}$. For simplicity we also restrict ourselves to $z=0$ and $\omega_x=\omega_y=0$, i.e. the spin of the particle must be perfectly aligned or anti-aligned with the orbital angular momentum. We know from the results of our numerical simulations that for moderate spins, these two extremes more or less bracket the particles with intermediate alignments.

In this 2D approximation, we can write out each component of the force,
\begin{eqnarray}
\label{eq:Fr}
\mathcal{F}_r & = & -\eta(2-\eta) \frac{G M_* m_p}{r^2} - F_D \frac{v_r}{v_{rel}}  \\
& & \ \ \ \ - F_L \frac{v_\phi}{v_{rel}} + 2 m_p \Omega_K (1-\eta) v_\phi \nonumber \\
\label{eq:Fth}
\mathcal{F}_\phi &=& - F_D \frac{v_\phi}{v_{rel}} +F_L \frac{v_r}{v_{rel}} -2 m_p \Omega_K (1-\eta) v_r \nonumber \\
& & \ \ \ \ + \frac32 m_p  \Omega_K v_r  \left(   1 - \frac{13}{21}\eta          \right)    
\end{eqnarray}
In this context, $v_{rel} = \sqrt{v_r^2+v_\phi^2}$. The factor of 2 in the first term of equation \ref{eq:Fr} arises from the fact that we are in a frame co-rotating with the gas, whose angular velocity is a factor of $1-\eta$ slower than Keplerian. This factor gets squared when evaluating the centrifugal force. The factor of $13/21$ in the final term comes from the $d\vec{\Omega}/dt$ term in the Euler force, and the resulting radial derivatives of $\Omega_K$ and $\eta$ following an application of the chain rule. 

Figure \ref{fig:stream} shows streamlines of this vector field as a function of $v_r$ and $v_\phi$ for various locations in the laminar disk model (different rows) for a large and a small value of the dimensionless spin $\mathcal{S}=s\omega/v_{rel}$ (left and right columns). The vector field for $\hat{\omega}_z>0$ is shown, while the field for $\hat{\omega}_z<0$ is not. Each panel of Figure \ref{fig:stream} has a red marker denoting $\mathcal{F}_r=\mathcal{F}_\phi=0$. The red dashed line indicates a circle of radius $\eta v_K$ centered on $v_r=v_\phi=0$. In the standard no-spin case, we would expect roots of $\mathcal{F}$ to lie near this line in the weakly-coupled regime, and as a particle entered the strongly-coupled regime the root would move towards zero velocity, which we do in fact see.

In addition to the red symbol, each panel of Figure \ref{fig:stream} includes a blue symbol, which is a root of the $\mathcal{F}$ vector field (not shown) when the spin is anti-aligned with the orbital angular momentum. In the low-spin case, we see that the red and blue points are virtually indistinguishable, i.e. the direction of the spin (and the spin itself) does not matter a great deal. As we expected from the numerical integrations, we see that in the high-spin case the zero-force velocities are appreciably different.

We posit that a population of spinning particles would have substantial velocity dispersions arising from two different effects visible in our numerical simulations. The difference in position between the red and blue points in each panel of Figure \ref{fig:stream} demonstrates that even if the particles are nearly in equilibrium (i.e. $\mathcal{F}_r=\mathcal{F_\phi}=0$), particles that are identical in every respect except their spin orientation will have different velocities. This explains the spread in time-averaged values of $v_{rel}$ at fixed radius visible in Figure \ref{fig:r1_avg}. 

The second effect arises from the fact that $\mathcal{F}_r \ne \mathcal{F}_\phi \ne 0$. This is quite clear in the numerical simulations, given the short-period oscillations in properties of the particles' trajectories (Figure \ref{fig:r1_raw}). These are the result of the continuous oscillations in $C_L \propto \hat{\omega} \times \hat{v}_{rel}$ that occur as particles with fixed spin-axis $\hat{\omega}$ (in the inertial frame) rotate about the star, and hence change $\hat{v}_{rel}$. This effect tends to be much smaller than the spread in velocities owing to the difference in equilibrium velocities between spin-aligned and anti-aligned particles. The exception is for particles whose spin axis lies nearly in the orbital plane, in which case $\hat{\omega} \times \hat{v}_{rel}$ has order unity oscillations. 

The streamlines in Figure \ref{fig:stream} are colored by a quantity similar in spirit to the dimensionless stopping time, namely $m_p \eta v_K \Omega_K/\mathcal{F}$. This is the ratio of the time it would take a particle to traverse a distance $\eta v_K$ in this space to the orbital time. This value tends to be of order unity or less, meaning if the particle is out of equilibrium it will move a substantial distance towards equilibrium in a single orbit. Moreover the streamlines lead more or less directly to the equilibrium point for the small-radii cases, indicating that the particle should be found close to its equilibrium point. If the Magnus force were stronger relative to the drag force, the streamlines would circle the equilibrium point many times, meaning that the particle could plausibly be frequently out of equilibrium, however this scenario seems unlikely based on equation \ref{eq:liftRatio}. 

Since we expect the particles to be near equilibrium, i.e. $\mathcal{F}_r = \mathcal{F}_\phi = 0$, we have a reliable way of estimating $v_r$ and $v_\phi$ for $\hat{\omega} = \pm \hat{z}$. Namely, we can numerically find the root of $\mathcal{F}(\vec{v})$ for arbitrary values of the particle size, location in the disk, and dimensionless spin. We denote these two equilibrium velocities $\vec{v}_+$, and $\vec{v}_-$. We posit that the difference in these two velocities is a reasonable proxy for the velocity dispersion of particles in the disk, to within a factor of a few. We find that the results of the numerical integrations are well-matched by the following crude estimate of the velocity dispersion 
\begin{equation}
\label{eq:sigmahat}
\hat{\sigma} \equiv \frac15 |\vec{v}_{+} - \vec{v}_{-}|.
\end{equation}
Figure \ref{fig:sigplot} shows $\hat{\sigma}/\eta v_K$ as a function of radius for several different spin values (left panel). When $\mathcal{S}=1$, we can directly compare this prediction to the standard deviation of $v_{rel}$ for our randomly-oriented sample of particles from the numerical simulations in the previous section. This $\sigma$ normalized by $\eta v_K$ is shown as the light blue points with error bars, and it indeed agrees reasonably well with our estimate $\hat{\sigma}$. Since it is not numerically difficult to calculate $\vec{v}_\pm$, equation \ref{eq:sigmahat} allows us to estimate the velocity dispersion for any particle size, spin rate, and heliocentric radius. It would also be straightforward to modify any ingredient of the fiducial model, e.g.  the mass of the star, the gas density, or the mean density of the particles.

\subsection{Gravitational and Streaming Instabilities}

\begin{figure*}
\centering
\hspace*{-.01cm}\includegraphics[width=15cm]{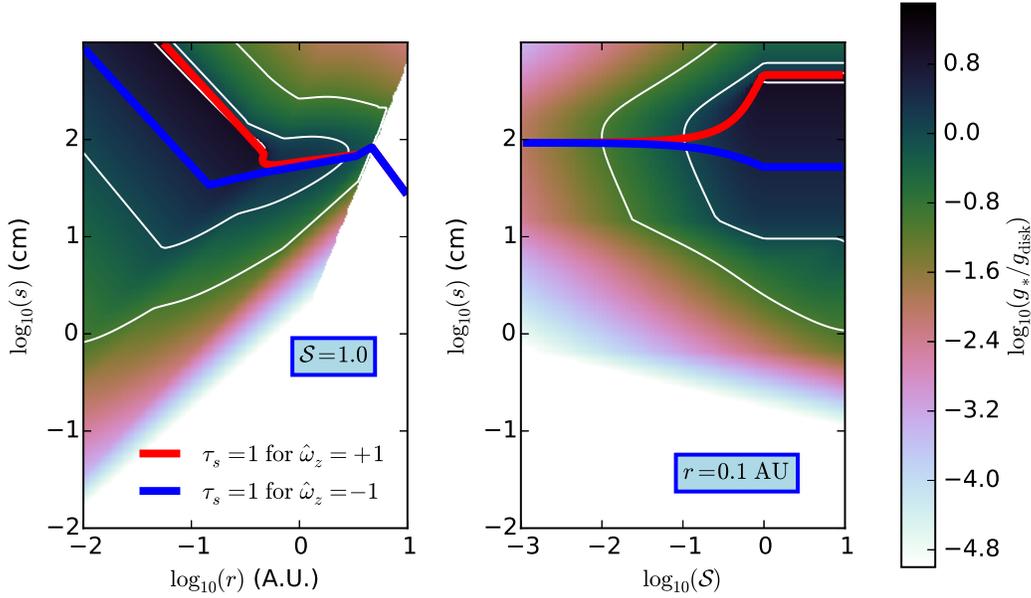}
\caption{Slices of the ratio of $g_*/g_\mathrm{disk}$, a criterion used for the stability of a dust disk to gravitational collapse. Values greater than 1 mean that the disk is stable to collapse. White contours show $g_*/g_\mathrm{disk}=0.1$, $1$, and $10$. While the red and blue lines show where in this parameter space $\tau_s=1$, depending on whether their spins are aligned or anti-aligned with the gas angular momentum. This figure assumes that the velocity dispersion of the dust is well-approximated by Equation \ref{eq:sigmahat}. To evaluate this equation, we also assume that $\rho_s=1\ {\rm g}\ {\rm cm}^{-3}$, $F=Z_{rel}=1$, and $M_*=1 M_\odot$. If monolithic gravitational collapse is relevant in disks, this figure shows that the Magnus force alone is strong enough to stop it for large enough dimensionless spin $\mathcal{S}$ and particle size $s$, and small enough heliocentric radius $r$. }
\label{fig:stabilityHeatmap}
\end{figure*}

\begin{figure*}
\centering
\hspace*{-.01cm}\includegraphics[width=15cm]{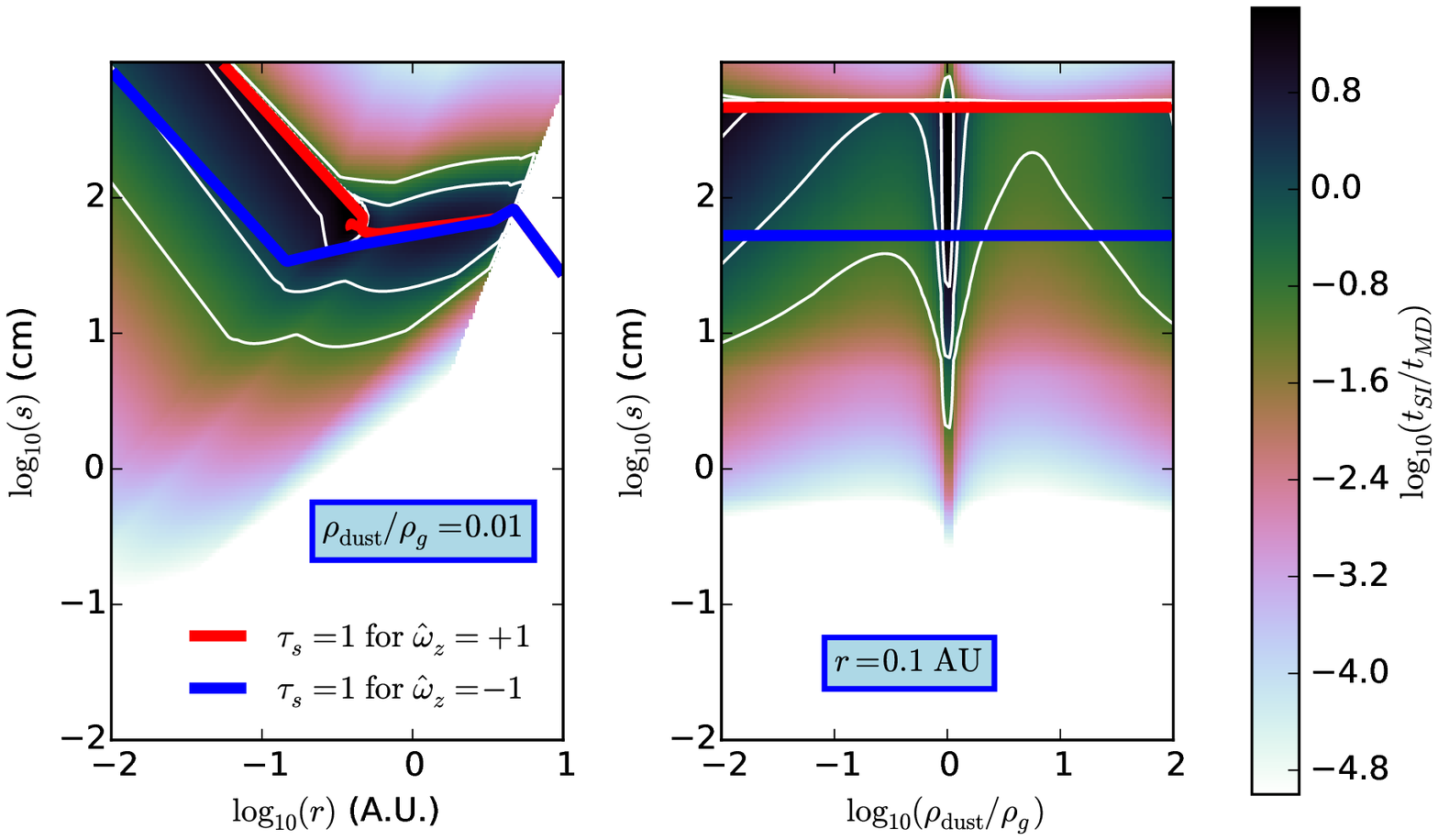}
\caption{Slices through parameter space of the streaming instability growth timescale $t_{SI}$ compared to the diffusion timescale owing to the Magnus force $t_{MD}$. White contours are shown at $t_{SI}/t_{MD}=0.1$, $1$, and $10$. Contours where $\tau_s=1$ are shown in red and blue for particles with spins aligned and anti-aligned with the gas angular momentum respectively. In both panels $\mathcal{S}=1$, and other parameters of the problem are kept fixed at their fiducial values (e.g. $F=Z_{rel}=1$). There are parts of the parameter space where $t_{SI}/t_{MD} \ga 1$, meaning that the fastest growing mode of the streaming instability may be suppressed by diffusion owing to the Magnus force.}
\label{fig:growthRateHeatmap}
\end{figure*}

The velocity dispersion of particles plays an important role in planetesimal formation. In a scenario like that envisioned by \citet{Goldreich1973}, the stability of the dust disk is determined by the Toomre $Q$ parameter, which is directly proportional to the velocity dispersion at a given heliocentric radius and particle surface density. $Q=\Omega \sigma/(\pi G \Sigma_s)$ is shown in the right panel of \ref{fig:sigplot}. Another common metric is the ratio of the vertical component of the star's gravity to the self-gravity of the particle disk \citep{Garaud2004}. This stability criterion is
\begin{equation}
\frac{g_*}{g_\mathrm{disk}} = \frac{\Omega_K^2}{4\pi G(\rho_\mathrm{dust}+\rho_g)}
\end{equation}
Note that $\rho_\mathrm{dust}$ is the mass density of solid material in a large fixed region, as opposed to $\rho_s$, the average density of a single grain. Once again using $\hat{\sigma}$ as a proxy for the velocity dispersion of solid grains, we can estimate $\rho_\mathrm{dust} \sim \Sigma_s \Omega_K/\hat{\sigma}$, while $\rho_g$ is taken to be the midplane density given in equation \ref{eq:rhog}.

The value of this stability parameter is mapped out in Figure \ref{fig:stabilityHeatmap} as a function of $r$ and $s$ at fixed $\mathcal{S}$, and as a function of $\mathcal{S}$ and $s$ at fixed $r$. Throughout both panels we have kept $\rho_{s,0} = F = Z_\mathrm{rel} = 1$ and $M_*=1M_\odot$. As in previous figures, this ratio quickly drops to zero at sufficiently large radii, where particles are subject to Epstein drag and we presume the Magnus force to be inoperative. Nonetheless, we see that for small enough heliocentric radii and large enough dimensionless spin rates, the particles have a sufficient velocity dispersion to keep $g_*/g_\mathrm{disk} > 1$, preventing runaway vertical gravitational collapse.

For our default choice of parameters, the critical size for a collection of spinning particles to avoid runaway collapse is of order $s=10\ \mathrm{cm}$. This is the same scale on which grain collisions cease to be effective at increasing the mass of particles \citep{Blum2008}. One could therefore imagine a scenario in which particles coagulate up to 10 cm size-scales, and these particles collapse via gravitational instability to form km-size planetesimals {\em only in regions where the Magnus force is irrelevant}. This would pose a problem because terrestrial planets are incredibly common at small heliocentric radii \citep[e.g.][]{Petigura2013}.

This problem has several plausible solutions. For instance, planetesimals might only form at large radii, with the resulting planets migrating through the disk to their presently-observed locations at small radii. Another possibility is that the collective gravitational collapse may occur when particles are appreciably smaller than $10\ \mathrm{cm}$, in which case the Magnus force is unlikely to be strong enough to prevent this collapse. This is plausible as long as the sticking efficiency of particles is small enough that by the time particles can settle to the mid plane, they have not already coagulated to sizes larger than $\sim 10\ \mathrm{cm}$ \citep{Goldreich1973, Garaud2004}. It is also possible that the typical spin of particles is simply not large enough for the Magnus force to be important at small radii.

While one or more of these possibilities is likely true, an even easier solution is to pass the buck. The laminar dust disk assumed in the gravitational collapse scenario is easily disrupted by even moderate turbulence in the gas disk, and even in laminar disks, the dust disk is subject to Kelvin Helmholz instabilities from vertical shear \citep{weidenschilling1980}. The most popular mechanism to concentrate the dust enough for it to collapse gravitationally is the streaming instability. \citet[][YG05 hereafter]{Youdin2005} showed that the two-fluid gas+dust equations of motion are unstable to axisymmetric perturbations, even in the absence of self-gravity and vertical stratification. If the particles are allowed to move with respect to the fluid, i.e. $\tau_s>0$, these instabilities grow and concentrate particles on timescales slower than the dynamical time, but quickly enough that the particles do not drift into the star. The ability of the Magnus force to affect planet formation therefore likely depends on its effect on the streaming instability.

We find that the Magnus force may indeed stabilize certain modes of this instability, though the region of parameter space where this is true may be fairly small. If the orientation of particle spins is approximately isotropic, we can estimate that the diffusion coefficient associated with the Magnus force is 
\begin{equation}
D \sim  \sigma L,
\end{equation}
where once again we take $\sigma \approx \hat{\sigma}$. $L$ is the effective mean free path of the particles. To estimate this, we need to make a strong assumption about how frequently a particle's spin is reoriented. If collisions are responsible for keeping the particles spinning, we expect that the mean free path will be the particle's velocity times the typical time between collisions. We also expect that $L<H$, the scale height of the particle disk. We therefore take
\begin{equation}
L= \min\left( \sigma t_c, H\right),
\end{equation}
with $t_c$ given by equation \ref{eq:tColl}.

Following section 3.2.2 of YG05, we adopt $H\sim \eta r$, and we estimate the diffusion time for modes of wavenumber $k$,
\begin{equation}
t_{MD} = \frac{4\pi^2}{k^2 D}.
\end{equation}
The subscript MD is meant to distinguish diffusion owing to the Magnus force from other sources, e.g. turbulence. The largest modes we expect to be physically relevant will be of order the particle scaleheight, in which case $k\approx 2\pi/H$, but substantially smaller modes tend to grow quickest. YG05 provide the following fit\footnote{The conditionals $f_g>1/2$ and $f_g<1/2$ appear to be erroneously reversed in Equation 32 of YG05.} to the transverse wavenumber that maximizes the growth rate of the streaming instability at fixed $\eta r k_z=1$
\begin{equation}
\label{eq:kx}
\eta r k_x = 
\begin{cases}
\left(2\tau_s f_g^3 \right)^{-1/2} & \ \ f_g>1/2 \\
\sqrt{2/\tau_s} f_g^{-0.4} & \ \ f_g<1/2.
\end{cases}
\end{equation}
Here $f_g=\rho_g/(\rho_g+\rho_\mathrm{dust})$ is the gas fraction. For the $\tau_s \ll 1$ limit in which the derivation is valid, the dimensionless wavenumbers $\eta r k_x$ can range from order unity to hundreds.

The corresponding growth rate of the streaming instability is approximately (equation 44 of YG05)
\begin{equation}
\label{eq:tSI}
1/t_{SI} = 4 f_p f_g^2(f_p-f_g)^2 \frac{(\eta r k_x)^4}{(\eta r k_z)^2} \tau_s^3 \Omega_K,
\end{equation}
where $f_p = 1-f_g$ is the particle density fraction of the 2-fluid system. Using $\eta r k_z = 1$ and $k_x$ as defined by equation \ref{eq:kx}, we can map the ratio of the streaming instability growth rate to the diffusion rate, $t_{MD}/t_{SI}$. The diffusion rate is evaluated at a wavenumber larger than $k=\sqrt{k_x^2 + k_z^2}$ by a factor of $2\pi$ to correct for the fact that the wavenumber corresponding to one scale height is $2\pi/(\eta r)$, not $1/(\eta r)$. Figure \ref{fig:growthRateHeatmap} shows this ratio for a pair of slices through parameter space. We caution that this ratio is only an order of magnitude estimate, since it is unclear precisely which wavenumbers to use, how to define the diffusion coefficient, and since the growth rate from equation \ref{eq:tSI} is derived from an approximate dispersion relation which is not applicable for $\tau_s$ near or greater than unity. Moreover there is some ambiguity about which $\tau_s$ to use, since it varies as a function of $\mathcal{S}$.

The only region of the parameter space where we find that the Magnus force acts more quickly to diffuse particles than the streaming instability acts to aggregate them occurs near $\tau_s = 1$, precisely where Equation \ref{eq:tSI} becomes unreliable. This is not a coincidence, as both processes have their largest effect near $\tau_s =1$. The fact that diffusion owing to the Magnus force is a strong enough effect that it is comparable in magnitude to the streaming instability in this critical regime suggests that further investigation in the non-linear regime is warranted.

Over the past 10 years, numerous numerical simulations have been performed to study the nonlinear growth of the streaming instability \citep[e.g.][]{youdin2007, Johansen2007ApJ, Johansen2007, Yang2014}. Even if sufficiently small or sufficiently slowly spinning dust grains are initially subject to the streaming instability, at some stage during the collapse and coagulation of the concentrated dust the Magnus force may become important. As the particles increase in size, their stopping times and Reynolds numbers may pass through the critical values of $\tau_s \sim 1$ and $Re \sim 800$. Indeed, within $r \la 1\ A.U.$ particles pass through these critical values simultaneously when the particles are roughly 1 meter in radius, leading to velocity dispersions of order $\eta v_K$. Moreover, the spin of particles in these clumps may be large, since collisions occur frequently. Whatever small rotation rate is present in the initial large-scale region will be amplified by the conservation of angular momentum during the collapse to smaller size scales. \citet{Nesvorny2010} showed that this mechanism could account for binaries in the Kuiper Belt, and we suggest that it could also lead to high spin rates for individual objects formed in this process.

\section{Summary}

We have introduced the Magnus force as a potentially important piece of microphysics for solid bodies in gaseous protoplanetary disks. Rotating particles, by an asymmetry in the wake they leave in the background fluid, experience a force roughly perpendicular to both their direction of relative motion and their spin axis. Just like the drag force, the Magnus force depends on the dimensionless parameters of the flow, namely the Reynolds number $Re$, the dimensionless spin $\mathcal{S} = s\omega/v_{rel}$, and the ratio of the gas mean free path to the particle size, $\lambda/s$. This dependence produces three (particle-size-dependent) regions in the disk.

In the outermost region, we expect that the Magnus force will be irrelevant because $\lambda \ga s$. As particles drift inward in the disk eventually the gas begins to behave as a fluid, but the Magnus force remains comparatively weak. The only exception is if $\mathcal{S} \gg 1$, which we consider improbable. In the inner disk when $Re \ga 800$, the Magnus force becomes comparable to drag if $\mathcal{S} \sim 1$.

Clearly a great deal depends on the spin rate. In the modern solar system, most of the available data on spin rates is for kilometer-sized objects likely held together by self-gravity and confined to moderate spin rates. The smallest objects for which data is available show a hint towards higher spin-rates, scaling roughly as $\omega \propto s^{-1}$, in which case $\mathcal{S} \propto v_{rel}^{-1}$. In protoplanetary disks, spins can be both induced and damped by hydrodynamic drag, leading to a median $\mathcal{S} \ga 10^{-3}$ with an extremely broad log-normal distribution. Meanwhile frequent collisions between particles of comparable size may keep a large population of particles spinning quickly near $\mathcal{S} \sim 1$. Notably the orientation of the particle spins in these scenarios is likely close to isotropic. The coagulation of large groups of particles via the streaming instability or gravitational collapse may also induce large spins during this critical phase where particles may traverse the meter barrier.

Presuming that at least modest spin rates can be sustained in the disk, we carried out direct numerical integrations of the equations of motion for individual meter-sized objects to understand their dynamics when the Magnus force is included. Even when the Magnus force is comparable to the drag force, and even when the orientation of the particle is assumed to remain constant throughout the particle's life, the effect on particle lifetimes in the disk is moderate, at most a factor of two. Qualitatively, the spinning particles behave quite similarly to the non-spinning particles, gradually spiraling in to the central star. The trajectory followed by the spinning particles is determined by $\hat{\omega}_z$, the component of the spin rate parallel to the gas angular momentum, with aligned particles drifting in faster, and anti-aligned particles surviving longer relative to their non-spinning counterparts. 

The Magnus force can induce small but non-negligible out-of-plane motions, and perhaps most surprisingly, it can reverse the azimuthal velocity of particles relative to the gas. Non-spinning particles always orbit more quickly than the pressure-supported gas, but particles whose spins are aligned with the gas angular momentum actually orbit more slowly than the gas. This can be understood geometrically in a frame co-rotating with the gas. When a particle enters the strongly-coupled ($\tau_s \la 1$) regime, its inward radial motion becomes larger relative to its azimuthal velocity, and so the drag force begins to point outward. Spinning particles of opposite alignment will have Magnus force vectors pointing $\pm 90^\circ$ away from the drag force vector. Particles with the Magnus force pointing in the negative azimuthal direction will then acquire negative equilibrium velocities in this frame, i.e. they will orbit the star more slowly than the gas!

At a fixed radius, the velocity of particles with different spins can vary by up to a factor of two. Although a factor of two is unlikely to be significant in terms of particle lifetimes in the disk (the meter barrier is a problem regardless of whether particles spiral inwards in 100 or 200 years), this spread in velocities is large enough to have implications for theories of planetesimal formation. We show, by analyzing the equilibrium velocities of particles with oppositely-aligned spins in a frame co-rotating with the gas, that the velocity dispersion of a population of spinning particles is large enough to prevent monolithic runaway gravitational collapse of the dust layer in the inner regions of the disk for particles larger than about 10 centimeters. This suggests that the planets found in abundance at small orbital periods in observations either formed at larger radii, or that the planetesimals which would go on to form these planets formed from dust grains smaller than 10 centimeters. If this is the case, the gravitational collapse that formed these planetesimals had to occur more quickly than the timescale for grains to coagulate to sizes larger than 10 centimeters. In other words, the particles are subject to a race between settling by vertical gas drag and growth by coagulation. Counterintuitively, if coagulation wins planetesimal formation may be suppressed by the diffusive effects of the Magnus force.  

Monolithic gravitational collapse can also be disrupted by larger-scale turbulence in the disk and shearing instabilities that tend to stir up very thin dust disks. Attention has therefore turned to other means of concentrating particles, especially the streaming instability. The two-fluid gas plus dust equations of motion with non-zero $\tau_s$ exhibit a growing axisymmetric mode which concentrates dust on reasonably short timescales. We compare the linear growth rate with the timescale on which particles diffuse owing to the Magnus force, assuming the spin axes of the particles are reoriented on a collisional timescale. We find that diffusion owing to the Magnus force is comparable in strength to concentration owing to the streaming instability when $\tau_s \sim 1$, although this result will require multi-dimensional simulations to verify. The Magnus force may play an even more important role in the nonlinear evolution of this instability since the collapse of large-scale modes will tend to spin up particles by conservation of angular momentum.

\label{sec:summary}

\section*{Acknowledgements}
This work was supported by the National Science Foundation Graduate Research Fellowship under Grant No. DGE1339067, and in part by a UC/Lab Fee grant. Integrations were performed in parallel on UCSC's Hyades supercomputer, supported by the NSF under Grant Number 1229745. I would like to thank Greg Laughlin and Doug Lin for their advice and encouragement, Konstantin Batygin for a helpful conversation, and the anonymous referee for their thorough and helpful reports.

\bibliography{zotlib}

\clearpage

\end{document}